\documentclass[preprints,article,accept,pdftex,moreauthors]{Definitions/mdpi} 

\firstpage{1} 
\pubvolume{1}
\issuenum{1}
\articlenumber{0}
\pubyear{2025}
\copyrightyear{2025}
\externaleditor{Firstname Lastname} 
\datereceived{30 April 2025} 
\daterevised{15 June 2025} 
\dateaccepted{21 June 2025} 
\datepublished{ } 
\hreflink{https://doi.org/} 

\newcommand{\aj}{{ Astron. J.}}

\newcommand{\apj}{{ Astrophys. J.}}
\newcommand{\apjl}{{ Astrophys. J. Lett.}}

\newcommand{\apss}{{Astrophys. Space Sci.}}
\newcommand{\aap}{{Astron. Astrophys.}}
\newcommand{\aapr}{{Astron. Astrophys. Rev.}}
\newcommand{\aaps}{{Astron. Astrophys. Suppl.}}

\newcommand{\mnras}{{Mon. Not. R. Astron. Soc.}}

\newcommand{\pasp}{{Publ. Astron. Soc. Pac.}}

\newcommand{\actaa}{{Acta Astron.}}

\usepackage{xspace}

\newcommand{\msun}{{{\rm M}_{\odot}}}

\usepackage{amsmath, bm}

\Title{Extraction of Physical Parameters of RRab Variables Using Neural Network Based Interpolator}

\TitleCitation{Extraction of Physical Parameters of RRab Variables Using Neural Network Based Interpolator}

\Author{{Nitesh} 
 Kumar $^{1, \dagger}$*\orcidB{}, Harinder P. Singh $^{2,\dagger}$\orcidA{}, Oleg Malkov $^{3,\dagger}$, Santosh Joshi $^{4}$, Kefeng Tan $^{5}$, Philippe Prugniel $^{6}$ and Anupam Bhardwaj$^{7}$\orcidC{}}

\AuthorNames{Nitesh Kumar, Harinder P. Singh, Oleg Malkov, Santosh Joshi, Kefeng Tan, Philippe Prugniel and Anupam Bhardwaj}

        \AuthorCitation{Kumar, N.; Singh, H.P.; Malkov, O.; Joshi, S.; Tan, K.; Prugniel, P.; Bhardwaj, A.}

\address{%
$^{1}$ \quad Department of Physics, Applied Science Cluster, University of Petroleum and Energy Studies (UPES), Dehradun - 248007, Uttarakhand, India\\ 
$^{2}$ \quad Department of Physics \& Astrophysics, University of Delhi, {Delhi - 110007,} India; hpsingh@physics.du.ac.in\\
$^{3}$ \quad Institute of Astronomy of the Russian Academy of Sciences (INASAN), 48 Pyatnitskaya St., Moscow 119017, Russia; malkov@inasan.ru\\
$^{4}$ \quad Aryabhatta Research Institute of Observational Sciences (ARIES), Manora Peak, Nainital - 263001, Uttarakhand, India; santosh@aries.res.in\\
$^{5}$ \quad National Astronomical Observatories, Chinese Academy of Sciences (NAOC), 20A Datun Road, Chaoyang District, Beijing 100101, China; tan@nao.cas.cn\\
$^{6}$ \quad Centre de Recherche Astrophysique de Lyon (CRAL), Observatoire de Lyon, 9 Avenue Charles André, 69230 Saint-Genis-Laval, France; prugniel@obs.univ-lyon1.fr\\
$^{7}$ \quad Inter-University Centre for Astronomy and Astrophysics (IUCAA), Post Bag 4, Ganeshkhind, Pune 411007, India; {anupam.bhardwaj@iucaa.in}
}

\corres{Correspondence: nitesh.kumar@ddn.upes.ac.in}

\firstnote{{These authors contributed equally to this work.}}

\abstract{
Determining the physical parameters of pulsating variable stars such as RR Lyrae is essential for understanding their internal structure, pulsation mechanisms, and evolutionary state. In this study, we present a machine learning framework that uses feedforward artificial neural networks (ANNs) to infer stellar parameters—mass ($M$), luminosity (log($L/L_\odot$)), effective temperature (log($T_{\rm eff}$)), and metallicity ($Z$)—directly from Transiting Exoplanet Survey Satellite 
(TESS) light curves. The network is trained on a synthetic grid of RRab light curves generated from hydrodynamical pulsation models spanning a broad range of physical parameters. We validate the model using synthetic self-inversion tests and demonstrate that the ANN accurately recovers the input parameters with minimal bias. We then apply the trained model to RRab stars observed by the TESS. The observed light curves are phase-folded, corrected for extinction, and passed through the ANN to derive physical parameters. Based on these results, we construct an empirical period–luminosity–metallicity (PLZ) relation: log($L/L_\odot$) = (1.458 $\pm$ 0.028) log($P$/days) + (–0.068 $\pm$ 0.007) [Fe/H] + (2.040 $\pm$ 0.007). This work shows that ANN-based light-curve inversion offers an alternative method for extracting stellar parameters from single-band photometry. The approach can be extended to other classes of pulsators such as Cepheids and Miras.}


\begin{document}
\makeatletter
\def\UrlAlphabet{%
      \do\a\do\b\do\c\do\d\do\e\do\f\do\g\do\h\do\i\do\j%
      \do\k\do\l\do\m\do\n\do\o\do\p\do\q\do\r\do\s\do\t%
      \do\u\do\v\do\w\do\x\do\y\do\z\do\A\do\B\do\C\do\D%
      \do\E\do\F\do\G\do\H\do\I\do\J\do\K\do\L\do\M\do\N%
      \do\O\do\P\do\Q\do\R\do\S\do\T\do\U\do\V\do\W\do\X%
      \do\Y\do\Z}
\def\UrlDigits{\do\1\do\2\do\3\do\4\do\5\do\6\do\7\do\8\do\9\do\0}
\g@addto@macro{\UrlBreaks}{\UrlOrds}
\g@addto@macro{\UrlBreaks}{\UrlAlphabet}
\g@addto@macro{\UrlBreaks}{\UrlDigits}
\makeatother
\section{Introduction}
RR Lyrae stars are low-mass (0.5 $\lesssim$ M/$\msun \lesssim$ 0.8), evolved stellar objects {(}age $\gtrsim$ 10~Gyr \citep{savino_a_age_2020}) 
that occupy the intersection of the horizontal branch and the classical \emph{instability strip} in the Hertzsprung–Russell diagram. These stars are undergoing a core helium-burning phase, which is similar to the evolutionary stage of intermediate-mass classical Cepheids \mbox{(3 $\lesssim$ M/$\msun \lesssim$ 13).} Owing to their well-established period–luminosity relations (PLRs) in the infrared regime—originally identified by \citet{longmore_rr_1986} and subsequently refined by several studies (e.g., \citep{bono_theoretical_2001, catelan_rr_2004-1, sollima_rr_2006, muraveva_new_2015, bhardwaj_rr_2021})—RR Lyrae variables serve as reliable distance indicators and are critical to calibrating the cosmic distance ladder \citep{beaton_carnegie-chicago_2016, bhardwaj_high-precision_2020}. Furthermore, these stars offer important insights into stellar evolution and pulsation physics \citep{catelan_horizontal_2009}, and~they are effective tracers of ancient stellar populations in various galactic environments \citep{kunder_impact_2018}.

The physical parameters of RR Lyrae stars are traditionally estimated using empirical relations that connect various light curve characteristics—such as Fourier decomposition parameters and color indices—with the pulsation period \citep{cacciari_multicolor_2005, deb_physical_2010, nemec_fourier_2011}. However, advances in the theoretical modeling of stellar pulsation have enabled the construction of extensive model grids to investigate the intrinsic properties of RR Lyrae and other variable \mbox{stars~\citep{marconi_new_2015, de_somma_extended_2020, de_somma_updated_2022}.} Notably, the~radial stellar pulsation (RSP) module developed by \citet{smolec_convective_2008}, implemented within the Modules for Experiments in Stellar Astrophysics framework (\texttt{MESA}; \citep{paxton_modules_2011, paxton_modules_2013, paxton_modules_2015, paxton_modules_2018, paxton_modules_2019}), offers a powerful tool for generating theoretical pulsation models. These models facilitate the derivation of fundamental stellar parameters—such as mass, luminosity, and~effective temperature—thereby providing deeper insights into the structure and evolution of RR Lyrae~stars.

The motivation for inferring fundamental physical parameters—such as mass, luminosity, effective temperature, and~metallicity—from light curves comes from the proliferation of time-domain photometric data from large-scale surveys. While double-lined eclipsing binaries and resolved spectroscopic binaries provide the most precise parameter estimates, such systems are rare with~only a few hundred thoroughly characterized examples \citep{torres_accurate_2010, southworth_debcat_2015}. In~contrast, wide-field photometric surveys like EROS \citep{grison_eros_1995}, MACHO~\citep{alcock_rr_1998}, OGLE \citep{rucinski_eclipsing_2001}, ASAS \citep{pojmanski_all_2002}, TrES \citep{alonso_tres-1_2004}, HAT \citep{bakos_wide-field_2004}, CoRoT \citep{loeillet_doppler_2008}, and~Kepler \citep{matijevic_kepler_2012} have yielded millions of high-quality light curves for pulsating variables, including RR Lyrae stars. This unprecedented data volume makes light curve-based inference methods increasingly attractive for characterizing stellar populations on a large scale, particularly where spectroscopic data are~lacking.

An alternative method for inferring the physical parameters of RR Lyrae stars involves directly comparing observed light curves with a reference library of theoretical models. For~instance, \citet{das_variation_2018} applied this approach to a sample of RRab stars in the Large Magellanic Cloud (LMC), estimating their physical parameters by matching observed light curves with those from the model grid of \citet{marconi_new_2015}. However, this technique is constrained by the limited coverage of the model grid, as~only a small subset of observed LMC light curves closely matched the available theoretical templates. A~similar technique is used by \citet{kumar_physical_2025} for deriving physical parameters of stars in a globular cluster using medium resolution~spectra. 

A more sophisticated strategy involves constructing a denser and smoother model grid and employing non-linear optimization techniques for parameter estimation \citep{bellinger_fundamental_2016}. Nevertheless, generating such a comprehensive model grid is computationally demanding, posing practical limitations. To~overcome this challenge, \citet{kumar_predicting_2023} developed an artificial neural network (ANN) trained on the model grid of \citet{marconi_new_2015} for RRab stars in the $V$ and $I$ photometric bands. The~resulting ANN\endnote{The RRab interpolator for generating RRab light curves for a given combination of physical parameters is publicly available at \url{https://ann-interpolator.web.app/} {(accessed on 22 June 2025).}} 
 serves as an efficient interpolator that is capable of producing high-resolution, smooth model light curves in approximately 55~ms per sample, thereby significantly accelerating the parameter inference process. Hence, we can increase the density of the models using this RRab interpolator. However, the~RRab interpolator reliably generates the light curves within the parameter space on which it was trained (see \citet{kumar_predicting_2023}).

Artificial neural networks (ANNs) offer several advantages in the modeling and inference of pulsating variable star light curves. Once trained, ANNs can perform parameter inference on observed light curves within milliseconds, which is significantly faster than traditional methods such as grid-based forward modeling or interpolation using physical templates \citep{bellinger_when_2020, kumar_predicting_2023, kumar_multiwavelength_2024}. Furthermore, ANNs allow the generation of a denser and smoother model grid across the parameter space, improving the fidelity and resolution of the inferred parameters. In~our case, the~ANN-based interpolator enables efficient mapping from physical parameters to light curves, and~vice~versa, over~a finely sampled synthetic grid. While the current work focuses on the $I$ band, which closely resembles the TESS passband, the~approach is generalizable to other photometric bands provided suitable training data are available. These benefits make ANNs a powerful alternative to classical interpolation schemes in the context of variable star~analysis.

In this study, we generated a smooth grid of model light curves in the \textit{I} band over a given parameter space and then trained a reverse interpolator as discussed in \citet{kumar_multiwavelength_2024} to obtain the physical parameters of non-Blazhko fundamental mode RR Lyrae stars observed in the Transiting Exoplanet Survey Satellite (TESS) field. The~\textit{I} band model light curves were preferred over the \textit{V} band for training the reverse interpolator, as~the TESS light curves exhibit greater similarity to the \textit{I} band, thereby enabling more accurate parameter estimation from TESS~observations.

\section{Data and Theoretical~Grid}\label{sec:data}
\unskip
\subsection{Synthetic Grid~Construction}
To accurately determine the physical parameters of pulsating variable stars, it is necessary to compare observed light curves with a grid of model light curves. However, pre-computed grids of hydrodynamical models are typically coarse and unevenly distributed across parameter space, owing to the high computational cost and time required to solve the time-dependent equations governing stellar pulsations. Furthermore, constraints on parameters such as mass, surface gravity, and~metallicity often rely on spectroscopic measurements, which are not always available for photometric datasets. Consequently, constructing a fine, dense grid of models becomes essential for reliably inferring stellar~properties.

In this work, we generated a fine synthetic grid of RRab light-curve templates in the $I$ band using the RRab interpolator trained by \citet{kumar_predicting_2023}. We adopted a finer and more uniform sampling where mass ($M$) varies from 0.5 to 0.8\,$M_\odot$ with a constant step size of 0.05\,$M_\odot$, luminosity ($\log(L/L_\odot)$) spans from 1.50 to 2.00\,dex with a step size of 0.0555\,dex, and~effective temperature ($T_{\mathrm{eff}}$) ranges from 5000\,K to 8000\,K with a step size of approximately 88.23\,K. The~metallicity ($Z$) covers the range from $10^{-4}$ to $10^{-2}$ in seven logarithmic steps. The~adopted parameter boundaries correspond to the limits of the original training parameter space of the interpolator, and~the step sizes were selected to ensure uniform coverage across each parameter dimension. For~each value of metallicity, the~hydrogen abundance ($X$) is computed using the relation $X = 1 - Y - Z$, assuming a fixed helium fraction $Y = 0.245$. The~parameter space of the synthetic grid is shown in Table~\ref{tab:grid_parameters}. Using the trained interpolator, we generate synthetic template light curves in the $I$ band for a grid comprising 17,150 distinct combinations of stellar~parameters.
\begin{table}[H]
    \centering
    \caption{The synthetic model light curves are generated in the $I$ band using the RRab interpolator for the given combination of the physical~{parameters.}} 
\begin{tabularx}{\textwidth}{CCC}
        \toprule
        \textbf{Parameter} & \textbf{Range} & \textbf{Step Size} \\
        \midrule
        Mass ($M/M_\odot$) & 0.5--0.8 & 0.05 \\
        $\log(L/L_\odot)$ & 1.50--2.00 & 0.0555 \\
        $T_{\mathrm{eff}}$ (K) & 5000--8000 & 88.23 \\
        Metallicity ($Z$) & $10^{-4}$--$10^{-2}$ & logarithmic steps \\
        Hydrogen fraction ($X$) & 0.755--0.745 & computed as $1 - Y - Z$ \\
        \bottomrule      
    \end{tabularx}
    \label{tab:grid_parameters}
\end{table}
The pulsation period ($P$) of an RR Lyrae star is closely linked to its mass, luminosity, and~effective temperature, as~originally described by the van Albada--Baker (vAB) relation~\cite{van_albada_masses_1971}. We employ a modern version of the vAB relation, incorporating the dependence on metallicity, as~formulated by \citet{marconi_new_2015}, to~compute the periods for our synthetic models. This relation is specifically calibrated for fundamental-mode (RRab) pulsators.

\subsection{Observational TESS~Sample}
The Transiting Exoplanet Survey Satellite (TESS) is a NASA Astrophysics Explorer mission employing four wide‐field CCD cameras to perform a nearly all‐sky photometric survey, providing a pre‐selected 2‐minute cadence and 30‐minute full‐frame images (FFIs) covering approximately 2300 deg$^{2}$ per sector~\cite{ricker_transiting_2015}. The~continuous, high‐precision optical time‐series photometry provided by TESS has revolutionized the study of pulsating variables by offering uniformly sampled, multi‐sector light curves free from diurnal gaps~\cite{ricker_transiting_2015}. Among~its many variable‐star discoveries, TESS has observed over a thousand RR Lyrae variables, including hundreds of fundamental‐mode RRab pulsators, allowing detailed analyses of their pulsation modes, the~incidence and characteristics of the Blazhko effect, and~low‐amplitude secondary oscillations~\cite{molnar_first_2021}. Differential‐image photometry techniques applied to TESS FFIs have cleanly extracted RRab light curves, which, when combined with Gaia parallaxes, have been used to refine empirical period–luminosity–metallicity relationships and to classify RR Lyrae populations across the sky~\cite{molnar_first_2021}.

\begin{figure}[H]
\includegraphics[width=10 cm]{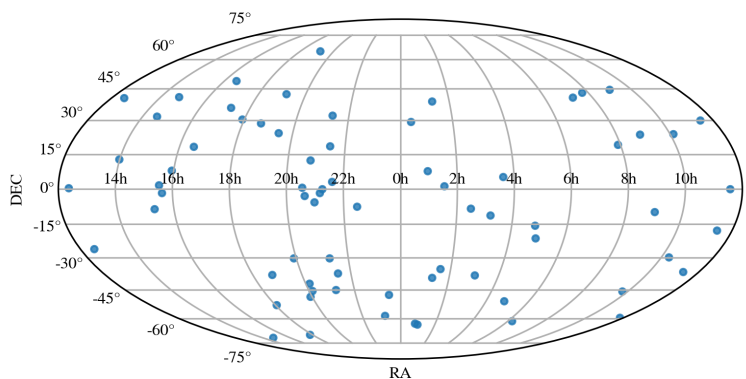}
\caption{{RA(J2000.0)-Dec(J2000.0)} 
The distribution of the 71 RRab stars cross-matched between SIMBAD and the TESS Input Catalog v8.2 \citep{2018AJ....156..102S, stassun_revised_2019, 2021arXiv210804778P}.}
\label{fig:RRab-distribution}
\end{figure}

We compiled a sample of RRab stars by cross-matching known RRab variables from the SIMBAD database with the TESS Input Catalog version 8.2 \citep{2018AJ....156..102S, stassun_revised_2019, 2021arXiv210804778P}. The~selection criteria required that each star had been observed in at least one TESS sector and had available measurements for at least one of the following parameters: effective temperature ($T_{\mathrm{eff}}$), surface gravity ($\log g$), or~metallicity ([Fe/H]). This process resulted in a final sample of 71 RRab stars. The~spatial distribution of these stars is shown in Figure~\ref{fig:RRab-distribution}. The~light curves of these stars were taken from the TESS archive using \textit{lightkurve}\endnote{\url{https://lightkurve.github.io/lightkurve/} {(accessed on 22 June 2025)}.} \cite{lightkurve_collaboration_lightkurve_2018} tool in Python. We converted the raw Pre-search Data Conditioning Simple Aperture Photometry (PDCSAP) flux to TESS magnitude using a zero point magnitude equal to $20.44$ from TESS data release notes\endnote{\url{https://archive.stsci.edu/files/live/sites/mast/files/home/missions-and-data/active-missions/tess/_documents/TESS_Instrument_Handbook_v0.1.pdf} {(accessed on 22 June 2025)}.}. We converted the TESS magnitude to I band magnitude by deriving a relation between the TESS magnitude and I band magnitude using the spectra of a typical RRab~star.

\unskip   

\section{Parameter Estimator ANN~model}

To derive the physical parameters of RRab stars observed by TESS, we first constructed a fine synthetic grid of model light curves covering a broad range of stellar masses, luminosities, effective temperatures, and~metallicities. The~template light curves were generated in the $I$ band, matching the photometric bandpass most closely related to the TESS instrumental response. The~synthetic grid comprised 17,150 distinct parameter combinations, each corresponding to a unique model light curve generated using the trained RRab interpolator. The~uniform and dense coverage of the grid enables robust interpolation across the relevant regions of parameter space and circumvents the limitations imposed by the sparsity of traditional hydrodynamical models. The~TESS light curves, corrected for interstellar extinction and folded on their best periods, serve as the observational inputs for the extraction of physical~parameters.

The relationship between the observed light curve ($y$) and the corresponding set of physical parameters ($\mathbf{x}$) can be expressed as a forward mapping function $f$ such that $y \equiv f(\mathbf{x})$. Inverting this mapping—recovering $\mathbf{x}$ given $y$—requires the construction of an inverse function $g$ such that $\mathbf{x} \equiv g(y) = f^{-1}(y)$. Artificial neural networks (ANNs) provide a powerful and flexible framework for approximating such inverse mappings, particularly when the underlying functions are continuous and differentiable \citep{cybenko_approximation_1989, hornik_multilayer_1989, hornik_approximation_1991}. By~training an ANN on the synthetic grid of light curves and associated stellar parameters, we enable the efficient and accurate retrieval of mass, luminosity, effective temperature, and~metallicity for observed RRab stars directly from their light-curve~morphology.

We trained a reverse interpolator using a synthetic grid of light curves and corresponding physical parameters. In~this setup, the~reverse interpolator is a feedforward ANN, where the input is the absolute $I$ band light curve, sampled at 500 equally spaced phase points between 0 and 1, and~the output consists of the physical parameters—mass ($M/M_{\odot}$), luminosity ($\log(L/L_{\odot})$), effective temperature ($\log(T_{\rm eff})$), and~metallicity ($Z$).

The input layer consists of 500 absolute $I$ band magnitude values, each corresponding to one of the sampled phase points. Since the physical parameters exhibit vastly different numerical ranges, we employed the \textit{Robust Scaler} method to scale the output values. This approach scales the data using the interquartile range (IQR), making the network less sensitive to outliers and leading to more stable and efficient training. By~transforming the physical parameter outputs into a more uniform scale, we ensured that the ANN training could converge more quickly and avoid issues stemming from disparities in the different~parameters.

The most crucial part in training any ANN is the choice of hyperparameters of the network, like the number of hidden layers, number of neurons in each hidden layer, activation function, optimization algorithm and learning rate. The~training time and the convergence of the ANN depend on these choices of hyperparameters. We used \texttt{RandomSearch} tuner from the \texttt{KerasTuner} library for conducting a systematic hyperparameter~tuning. 

We defined the model architecture in such a way that allows the number of hidden layers to vary between one and six. For~each hidden layer, the~number of units and activation function were treated as tunable parameters, and~the output layer was kept without any activation to predict the physical parameters. We chose the \texttt{\texttt{adam}} \citep{kingma_adam_2014} optimizer with the learning rate sampled logarithmically between $10^{-4}$ and $10^{-2}$. All layers were initialized with the \texttt{Glorot uniform} initializer. We trained each model for 50 epochs with a batch size of 128, using the mean squared error (MSE) as both the loss function and tuning~objective. 

The hyperparameter search explored 100 different configurations in total. Table~\ref{tab:hyperparams} summarizes the hyperparameter grid. The~dataset was divided into three parts: 80\% for training, 10\% for validation, and~the remaining 10\% for testing. The~final model selection was based on the configuration achieving the minimum validation~MSE.

\begin{figure}[H]
\includegraphics[width=\textwidth]{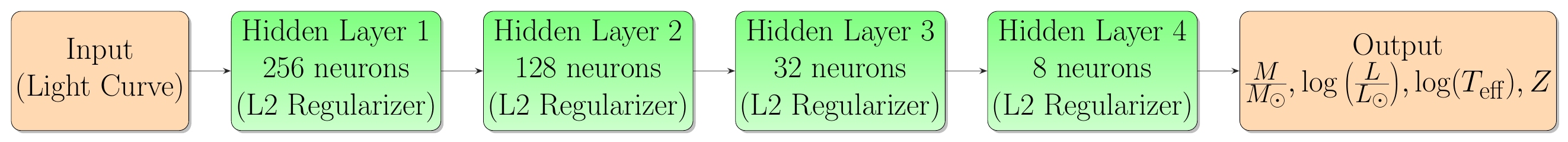}
\caption{\textls[-25]{ANN architecture adopted after hyperparameter optimization for physical parameter~estimation.}}
\label{fig:parameter_estimator}
\end{figure}

We achieved the lowest validation loss with an ANN model comprising $four$ hidden layers containing $256$, $128$, $32$, and~$8$ neurons, respectively. Each hidden layer has the \texttt{ReLU} activation function and includes L2 regularization with a strength of $10^{-6}$ to mitigate overfitting. The~output layer consists of four neurons and no activation function, corresponding to the four scaled predicted physical parameters. A~schematic diagram of the final ANN architecture is shown in Figure~\ref{fig:parameter_estimator}.
\begin{table}[H] 
\caption{Hyperparameter grid for ANN model~tuning.} 
\label{tab:hyperparams} 
\begin{tabularx}{\textwidth}{LL} 
        \toprule 
        \textbf{Hyperparameter} & \textbf{Values} \\
        \midrule 
        Number of layers & 1 to 6 \\
        Units per layer & 16, 32, 64, 128, 256, 512 \\
        Activation function & ReLU, tanh \\
        Learning rate & $10^{-4}$ to $10^{-2}$ (log-uniform sampling) \\
        Batch size & 32 (fixed) \\
        Optimizer & \texttt{adam} \citep{kingma_adam_2014} \\
        Kernel initializer & Glorot Uniform (fixed seed) \\
        Loss function & Mean Squared Error (MSE) \\
        \bottomrule 
    \end{tabularx} 
\end{table}\vspace{-9pt}

The final ANN model was trained using the \texttt{adam} optimizer with an initial learning rate of $0.001$ and~MSE as the loss function with~80\% of the original models used for training, 10\% for validation, and~the remaining 10\% reserved for testing. To~facilitate efficient convergence, we implemented a piecewise constant learning rate schedule: the learning rate was decreased by a factor of $5$ at batch sizes of $10$ and $100$, respectively. Training was performed over $1200$ epochs with a batch size of $32$. The~model weights were updated using shuffling at each epoch, and~training was parallelized across $eight$ CPU workers with multiprocessing. The~total training time was \textasciitilde $27.35$ min, utilizing a system equipped with $64$ CPU cores operating at a maximum frequency of $3.5$~GHz.

\begin{figure}[H]
    \includegraphics[width=7 cm]{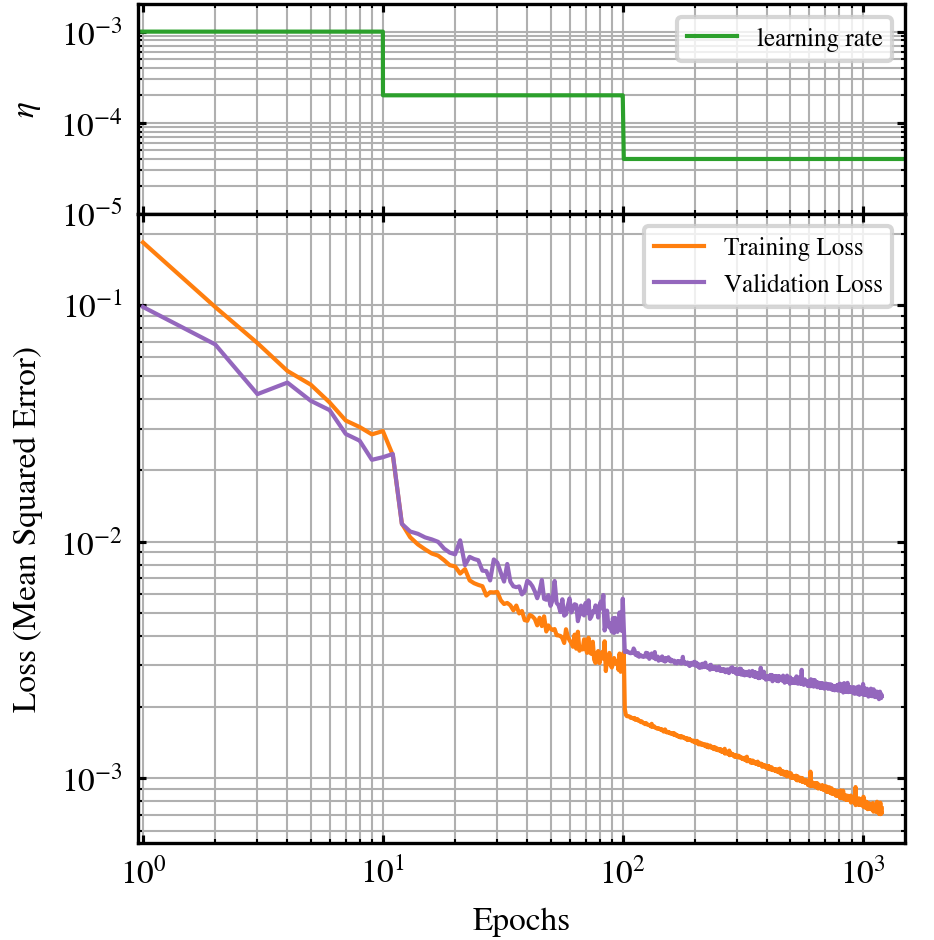}
    \caption{Training and validation loss curves, along with the learning rate progression, for~the ANN model over the training~epochs.}
    \label{fig:training_curve}
\end{figure}

The training and validation loss curve for the ANN model, along with the learning rate, is shown in Figure~\ref{fig:training_curve}. The~curve shows the training and validation loss (MSE) during the training over the course of 1200 epochs while also showing the variation in the learning rate. This figure provides insight into both the convergence behavior of the model and the effect of the learning~rate.

\section{Results}
\unskip
\subsection{Self~Inversion}

To assess the performance of the trained parameter estimator ANN model, we performed a self-inversion test. In~this test, the~synthetic model I band light curves from the test grid were passed through the trained ANN, and~the recovered physical parameters were compared with their original~values.





\begin{figure}[H]
    \includegraphics[width=\textwidth]{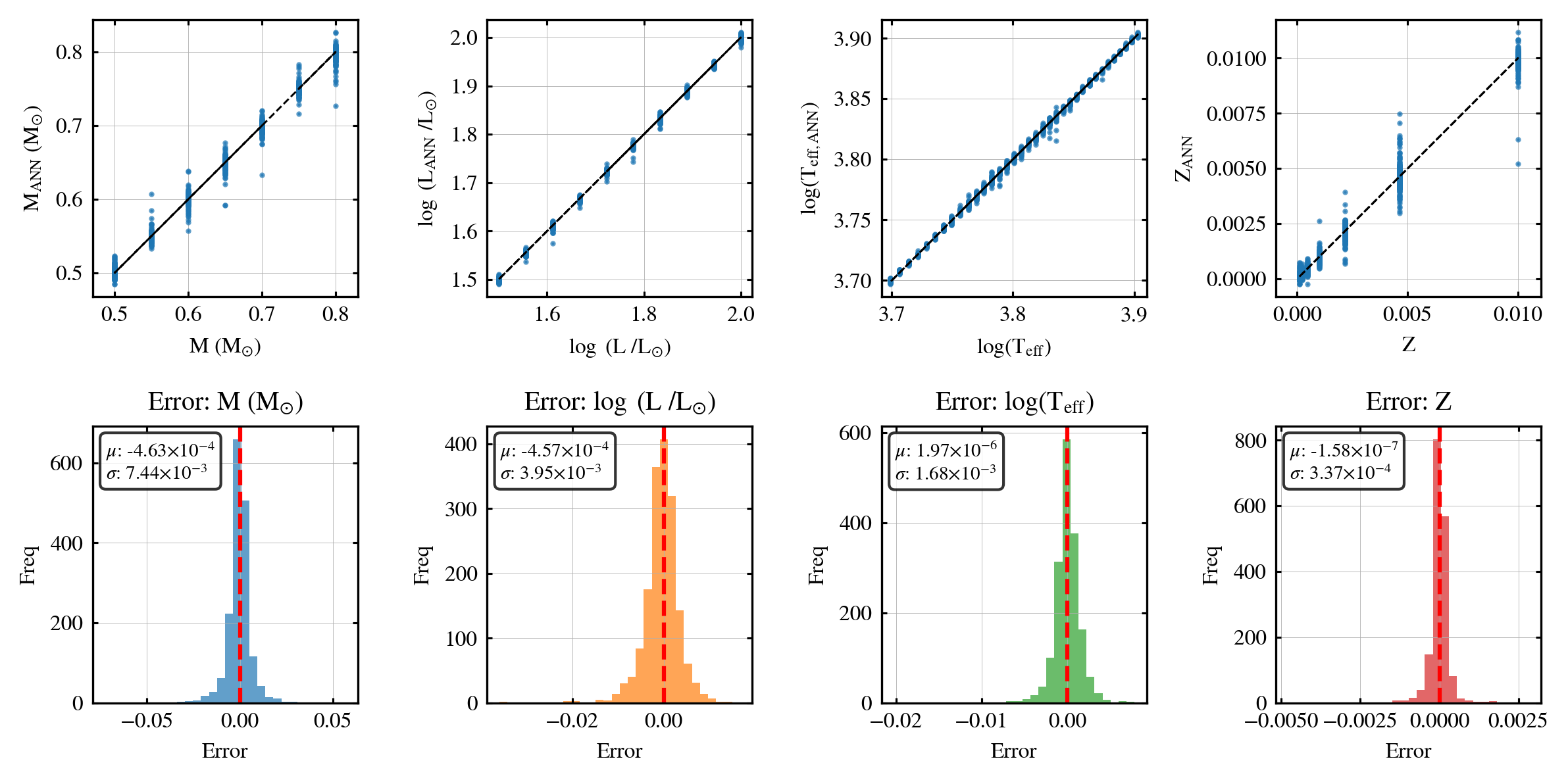}
    
    \caption{{Comparison} 
 between the original and predicted values of mass, $\log(L/L_\odot)$, $\log(T_{\rm eff})$ and Z from the synthetic self-inversion on the test set is shown in top panel. In~the bottom panel, the~error distribution is shown along with the mean and standard deviation of the error. The~close overlap indicates excellent recovery performance of the reverse interpolator for mass, luminosity and~temperature.}
    \label{fig:self_inversion_logL_logTeff}
\end{figure}

The comparison between the ANN-predicted and true values of mass, luminosity, effective temperature, and~metal abundance for the test set is shown in Figure~\ref{fig:self_inversion_logL_logTeff}. The~strong correlation between the predicted and true values demonstrates that the network accurately recovers stellar parameters across the entire grid. The~predicted points closely match the original values with~negligible scatter or systematic bias, as presented in Table~\ref{tab:performance_metrics}.

The relative root-mean-square error (Relative RMSE) provides an intuitive measure of a model's predictive performance across parameters with different numerical scales. It is defined as the root-mean-square error normalized by the mean absolute value of the true parameter values and is expressed as a percentage:

\[
\text{Relative RMSE} = \left( \frac{\sqrt{\frac{1}{n} \sum_{i=1}^n (x_i - \hat{x}_i)^2}}{\frac{1}{n} \sum_{i=1}^n |x_i|} \right) \times 100\%
\]
\noindent where \( x_i \) and \( \hat{x}_i \) are the true and predicted values of the parameter, respectively.

\begin{table}[H]
    \centering
    \caption{Performance of the trained parameter estimator model on the test set based on MSE and relative root-mean-square error (Relative RMSE). The~Relative RMSE is computed as the RMSE normalized by the mean absolute value of the true parameter~values.}
    \label{tab:performance_metrics}
\begin{tabularx}{\textwidth}{LCC}
        \toprule
        \textbf{Parameter} & \textbf{MSE} & \textbf{Relative RMSE (\%)} \\
        \midrule
        Mass ($M$) & $5.6 \times 10^{-5}$  & 1.15 \\
        Luminosity ($L$) & $1.6 \times 10^{-5}$ & 0.23 \\
        $\log(T_{\rm eff})$ & $3.1 \times 10^{-6}$  & 0.04 \\
        Metallicity ($Z$) & $1.1 \times 10^{-7}$ & 12.81 \\
        \bottomrule
    \end{tabularx}
\end{table}


The synthetic self-inversion tests confirm that the trained parameter estimator model is capable of recovering the physical parameters of RRab stars from their light curves with high accuracy and minimal bias. For~example, the~model achieves relative RMSE values of only 1.15\% for mass, 0.23\% for luminosity, and~0.04\% for $\log(T_{\rm eff})$, indicating excellent agreement between predicted and true values. This test validates the internal consistency and predictive reliability of the trained parameter estimator model, although~a relatively higher relative RMSE of 12.81\% is observed for~metallicity.

\subsection{Application to TESS RRab~Stars}

\subsubsection{Light-Curve Processing and~Folding}

To apply the trained parameter estimator model to real data, we selected a sample of RRab stars observed by the TESS for which literature estimates of fundamental stellar parameters are available in \citet{stassun_revised_2019}. The~TESS light curves were downloaded in the form of short-cadence (2-min) simple aperture photometry (SAP) flux from SPOC \citet{jenkins_tess_2016} pipeline FITS files. The~fluxes were converted into apparent magnitudes using the standard TESS zero point (adopted from TESS data release notes\endnote{{See note 3 above.}}): 
\begin{equation}
    m_{\rm TESS} = -2.5 \log_{10}(\text{flux}) + 20.44.
\end{equation}

The period for each light curve was derived using the Lomb--Scargle~\cite{lomb_least-squares_1976, scargle_studies_1982} method and then phase-folded using this derived period. Only positive flux points were retained to avoid contamination. To~suppress observational noise and fill missing data segments, we fitted a Fourier series to the folded light curve with 10 Fourier components (see Equation~(6) of \citet{kumar_multiwavelength_2024}). The~smoothed light curve was sampled at 1000 phase points and then rebinned to 500 evenly spaced bins between phases 0 and 1. This format matches the structure of the input layer used during ANN~training.

\subsubsection{Photometric Corrections and~Calibration}

To convert apparent TESS band magnitudes into absolute $I$ band magnitudes, several corrections were applied. Gaia DR3 parallaxes were used to compute distances, from~which the distance modulus $\mu$ was derived as
\begin{equation}
    \mu = 5 \log_{10}(d\, [{\rm pc}]) - 5.
\end{equation}

Extinction values were obtained from the IRSA Galactic Dust Reddening Tool, which returns extinction in $V$ band magnitudes ($A_V$) based on the star's equatorial coordinates using the reddening maps of the \citet{schlafly_measuring_2011}. The~$I$ band extinction $A_I$ was estimated using standard extinction ratios and~applied to the apparent TESS magnitudes after converting them to the $I$ band. As~TESS does not operate in the standard $I$ band, we employed the empirical transformation:
\begin{equation}
    I = m_{\rm TESS} - 0.0695.
\end{equation}

Thus, the~absolute $I$ band magnitude for each star becomes
\begin{equation}
    M_I = I - \mu - A_I.
\end{equation}

\subsubsection{ANN-Based Parameter~Inference}

The preprocessed and rebinned absolute $I$ band light curves were then passed through the trained ANN to predict the underlying stellar parameters. The~ANN output includes the stellar mass $M/M_{\odot}$, logarithmic luminosity $\log(L/L_\odot)$, effective temperature $\log(T_{\rm eff})$, and~metallicity $Z$:
\begin{equation}
    \left( M, \log L, \log T_{\rm eff}, Z \right) = \text{ANN}(y_{\text{TESS}}).
\end{equation}

The effective temperature can be recovered via inverse logarithmic transformation: $T_{\rm eff} = 10^{\log T_{\rm eff}}$. To~derive the iron abundance, we converted $Z$ to [Fe/H] using the relation provided by~\cite{piersanti_l_method_2007}
\begin{equation}
    [\mathrm{Fe/H}] = \log_{10}\left( \frac{Z/X}{Z_\odot/X_\odot} \right),
\end{equation}

\noindent where $Z_\odot = 0.0122$ and $X_\odot = 0.7392$ \citep{asplund_solar_2005}. The~hydrogen abundance $X$ is calculated from $X = 1 - Y - Z$, assuming a fixed helium mass fraction $Y = 0.245$. This value corresponds to the primordial helium abundance, which is suitable for RR Lyrae stars given their old evolutionary status. Helium enrichment in such populations is expected to be minimal. Observational support for this assumption is provided by \citet{marconi_gauging_2018}, who found little evidence of helium enhancement among RR Lyrae stars in the Galactic~bulge.


Table~\ref{tab:rrab_parameters} presents the derived physical parameters for a sample of RRab stars based on their TESS light curves, using our trained ANN-based parameter estimator. For~each star, the~table lists the TESS ID, parallax from the TESS Input Catalog v8.2 \citep{stassun_revised_2019}, the~literature period, and~the Lomb--Scargle period measured from the TESS data. The~effective temperature ($T_{\mathrm{eff}}$), mass ($M$), luminosity ($L$), and~metallicity ($Z$) inferred by the ANN are compared against corresponding values from the literature when available. We also report the photometric surface gravity ($\log g$).

\begin{figure}[H]
    \includegraphics[width=\linewidth]{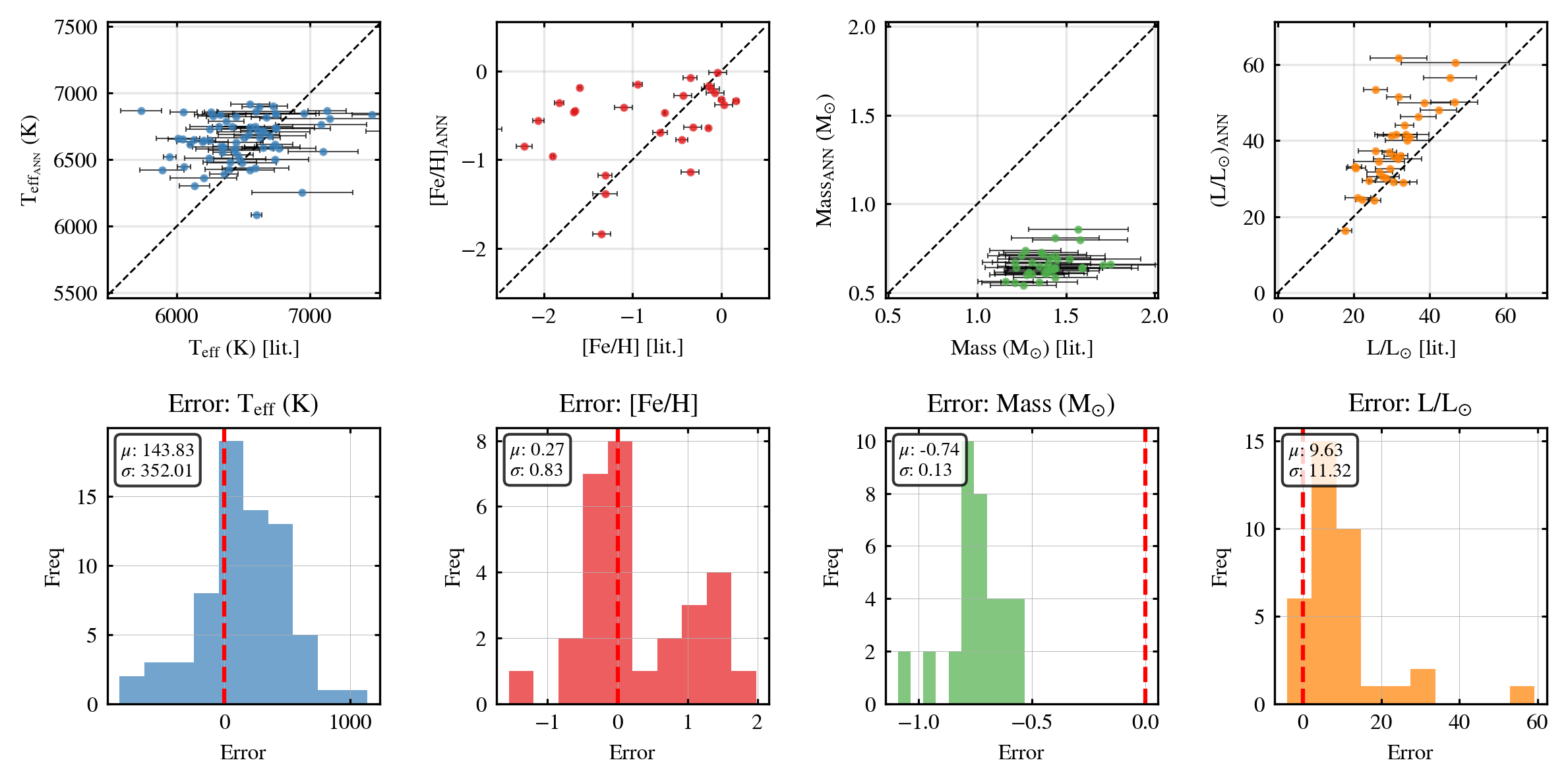}
    \caption{Comparison between literature values and ANN predictions for RR Lyrae stars. \textbf{Top row}: One-to-one comparison of literature values (x-axis) versus ANN predictions (y-axis) for effective temperature, metallicity, mass, and~luminosity with~error bars representing literature uncertainties. \textbf{Bottom row}: Distribution of prediction errors (ANN--literature) for each parameter, showing mean ($\mu$) and standard deviation ($\sigma$) values. The~dashed lines in all panels represent perfect~agreement.}
    \label{fig:param_comparison}
\end{figure}

Figure~\ref{fig:param_comparison} shows the comparison between the physical parameters of RR Lyrae stars derived from ANN analysis of TESS light curves and those from literature values. The~top row displays one-to-one comparisons with literature values on the x-axis and ANN predictions on the y-axis, while the bottom row shows error distributions. The~ANN method produces effective temperatures with a mean error of $143$\, K ($\sigma = 352$\, K), metallicities with a mean error of $0.27$\, dex ($\sigma = 0.83$\, dex), masses with a mean error of $-0.74\, M_{\odot}$ ($\sigma = 0.13\, M_{\odot}$), and~luminosities with a mean error of $9.63\, L_{\odot}$ ($\sigma = 11.32\, L_{\odot}$).

\startlandscape
\begin{table}[H]\small
\caption{Physical parameters of RR Lyrae stars derived from TESS light curves using ANN method compared with literature~values.}
\label{tab:rrab_parameters_1}
\begin{tabularx}{\textwidth}{lLCCCCCCCcCcCC}
\toprule
\textbf{ID} & \textbf{TESS ID} & \boldmath\textbf{Plx$_{\rm lit}$} & \textbf{Period} & \textbf{Period LS} & \boldmath\textbf{T$_{\rm eff_{lit}}$} & \boldmath\textbf{T$_{\rm eff_{ANN}}$} & \boldmath\textbf{M$_{\rm lit}$} & \boldmath\textbf{M$_{\rm ANN}$} & \boldmath\textbf{Lum$_{\rm lit}$} & \boldmath\textbf{Lum$_{\rm ANN}$ }& \boldmath\textbf{[Fe/H]$_{\rm lit}$} & \boldmath\textbf{[Fe/H]$_{\rm ANN}$} & \boldmath\textbf{logg$_{\rm lit}$} \\
 &  & \textbf{('') }& \textbf{(d)} & \textbf{(d)} & \textbf{(K)} & \textbf{(K)} & \boldmath\textbf{(M$_{\odot}$)} & \boldmath\textbf{(M$_{\odot}$)} & \boldmath\textbf{(L$_{\odot}$)} & \boldmath\textbf{(L$_{\odot}$)} & \textbf{(dex)} & \textbf{(dex)} & \textbf{(dex)}\\
\midrule
XZ Dra & 229913521 & 1.30 ± 0.02 & 0.476475 & 0.476574 & 6447 ± 224 & 6824 & 1.31 ± 0.22 & 0.67 & 34.76 ± 1.98 & 40.91 & -- & $-$0.56 & 3.21 ± 0.12 \\
V1195 Her & 232570970 & 0.53 ± 0.02 & 0.609729 & 0.608373 & 5889 ± 167 & 6420 & -- & 0.67 & -- & 56.91 & $-$1.60 ± 0.01 & $-$0.19 & -- \\
TW Lyn & 241174387 & 0.62 ± 0.04 & 0.481854 & 0.481025 & 6008 ± 162 & 6659 & -- & 0.62 & -- & 39.10 & $-$0.64 ± 0.01 & $-$0.47 & -- \\
TT Lyn & 29172806 & 1.22 ± 0.04 & 0.597443 & 0.597684 & 6191 ± 193 & 6635 & -- & 0.65 & -- & 52.51 & $-$1.65 ± 0.02 & $-$0.45 & -- \\
TW Boo & 68076619 & 0.72 ± 0.02 & 0.532264 & 0.531162 & 6247 ± 42 & 6645 & -- & 0.70 & -- & 45.32 & $-$1.46 ± 0.07 & -- & -- \\
SV Eri & 9843991 & 1.22 ± 0.06 & 0.713900 & 0.713374 & 6200 ± 249 & 6360 & -- & 0.81 & -- & 60.43 & $-$2.06 ± 0.06 &$-$0.56 & -- \\
SS Tau & 311091712 & 0.67 ± 0.05 & 0.369910 & 0.369768 & 6724 ± 108 & 6903 & 1.43 ± 0.26 & 0.64 & 21.14 ± 3.31 & 25.00 & 0.17 ± 0.03 & $-$0.33 & 3.53 ± 0.11 \\
RX Eri & 114923989 & 1.62 ± 0.03 & 0.587252 & 0.586150 & 6443 ± 205 & 6629 & -- & 0.63 & -- & 49.70 & -- & $-$0.56 & -- \\
U Lep & 146324929 & 0.93 ± 0.03 & 0.581458 & 0.580807 & 6541 ± 229 & 6689 & -- & 0.74 & -- & 58.66 & -- & -- & -- \\
TZ Aur & 328588068 & 0.66 ± 0.04 & 0.391673 & 0.391437 & 7128 ± 145 & 6867 & 1.59 ± 0.28 & 0.64 & 26.79 ± 3.39 & 31.72 & $-$0.15 ± 0.02 & $-$0.64 & 3.58 ± 0.10 \\
SZ Gem & 63172763 & 0.59 ± 0.04 & 0.501165 & 0.501342 & 6050 ± 101 & 6856 & -- & 0.71 & -- & 42.36 & $-$1.65 ± 0.13 & -- & -- \\
EZ Cnc & 197217727 & 0.50 ± 0.05 & 0.545777 & 0.546779 & 6625 ± 100 & 6662 & 1.39 ± 0.23 & 0.61 & 31.60 ± 6.34 & 35.19 & $-$0.00 ± 0.03 & $-$0.32 & 3.32 ± 0.13 \\
SS Gru & 129710422 & 0.14 ± 0.04 & 0.959791 & 0.956326 & 6942 ± 379 & 6252 & -- & 0.95 & -- & 291.22 & -- & $-$0.56 & -- \\
VW Scl & 41833926 & 0.85 ± 0.07 & 0.510917 & 0.511665 & 6672 ± 111 & 6816 & 1.41 ± 0.23 & 0.70 & 31.06 ± 5.55 & 41.71 & $-$1.27 ± 0.05 & -- & 3.35 ± 0.11 \\
Z Mic & 89358641 & 0.81 ± 0.07 & 0.586928 & 0.585981 & 6398 ± 200 & 6480 & 1.28 ± 0.21 & 0.60 & 34.01 ± 5.69 & 40.01 & -- & $-$0.18 & 3.19 ± 0.13 \\
VX Scl & 32282599 & 0.43 ± 0.03 & 0.637078 & 0.635441 & 6716 ± 208 & 6597 & -- & 0.76 & -- & 59.59 & -- & $-$0.95 & -- \\
RZ Cet & 1129237 & 0.68 ± 0.04 & 0.510600 & 0.511475 & 6248 ± 182 & 6729 & 1.21 ± 0.18 & 0.67 & 25.68 ± 3.12 & 53.38 & $-$1.90 ± 0.02 & $-$0.96 & 3.25 ± 0.12 \\
BN Aqr & 39999072 & 0.38 ± 0.06 & 0.469600 & 0.468717 & 6955 ± 760 & 6849 & 1.52 ± 0.40 & 0.69 & 46.69 ± 14.23 & 60.60 & -- & -- & 3.27 ± 0.31 \\
CP Aqr & 248483300 & 0.78 ± 0.09 & 0.463400 & 0.463006 & 7148 ± 522 & 6806 & 1.59 ± 0.32 & 0.64 & 26.65 ± 6.66 & 34.45 & -- & $-$0.74 & 3.59 ± 0.22 \\
SW Aqr & 387498678 & 0.76 ± 0.23 & 0.459305 & 0.459871 & -- & 6824 & -- & 0.72 & -- & 43.47 & -- & -- & -- \\
RR Cet & 344299442 & 1.52 ± 0.08 & 0.553041 & 0.552368 & 6650 ± 139 & 6589 & 1.40 ± 0.24 & 0.66 & 42.32 ± 4.75 & 48.09 & $-$1.35 ± 0.10 & $-$1.84 & 3.20 ± 0.10 \\
SX Aqr & 353029655 & 0.62 ± 0.07 & 0.535708 & 0.536120 & 6325 ± 194 & 6838 & 1.25 ± 0.19 & 0.71 & 31.85 ± 7.46 & 61.80 & -- & $-$1.84 & 3.19 ± 0.14 \\
AO Peg & 283289876 & 0.35 ± 0.04 & 0.547245 & 0.546917 & 6342 ± 71 & 6554 & -- & 0.67 & -- & 51.16 & $-$1.25 ± 0.07 & $-$2.81 & -- \\
SW And & 437761208 & 1.78 ± 0.16 & 0.442261 & 0.441881 & 6735 ± 138 & 6613 & 1.44 ± 0.24 & 0.59 & 30.35 ± 6.28 & 29.13 & $-$0.07 ± 0.10 & $-$0.25 & 3.38 ± 0.13 \\
DM Cyg & 117638854 & 0.97 ± 0.05 & 0.419868 & 0.419951 & 6415 ± 151 & 6748 & 1.29 ± 0.19 & 0.61 & 20.62 ± 2.50 & 32.85 & 0.03 ± 0.10 & $-$0.38 & 3.42 ± 0.10 \\
XX And & 186452465 & 0.69 ± 0.05 & 0.722772 & 0.721256 & 6097 ± 9 & 6613 & -- & 0.70 & -- & 67.68 & $-$1.67 ± 0.01 & $-$0.46 & -- \\
DH Hya & 47291018 & 0.47 ± 0.04 & 0.488996 & 0.488922 & 6259 ± 54 & 6856 & -- & 0.71 & -- & 59.94 & $-$1.52 ± 0.09 & -- & -- \\
RR Leo & 3941985 & 1.00 ± 0.09 & 0.452403 & 0.451699 & 6593 ± 291 & 6863 & 1.37 ± 0.25 & 0.72 & 33.79 ± 6.00 & 41.59 & -- & -- & 3.28 ± 0.16 \\
WY Ant & 168276785 & 0.91 ± 0.06 & 0.574350 & 0.575440 & 6765 ± 277 & 6583 & 1.45 ± 0.27 & 0.69 & 46.47 ± 6.23 & 50.15 & -- & -- & 3.21 ± 0.15 \\
V595 Cen & 152231997 & 0.44 ± 0.03 & 0.690994 & 0.691325 & 6430 ± 191 & 6569 & -- & 0.65 & -- & 53.30 & -- & $-$0.25 & -- \\
TU UMa & 144376546 & 1.56 ± 0.06 & 0.557658 & 0.556678 & 6200 ± 65 & 6643 & -- & 0.65 & -- & 44.82 & $-$1.31 ± 0.14 & $-$1.39 & -- \\
SS Leo & 49417864 & 1.13 ± 0.72 & 0.626351 & 0.626993 & -- & 6676 & -- & 0.70 & -- & 58.65 & -- & $-$1.80 & -- \\
W Crt & 219249305 & 0.75 ± 0.04 & 0.412012 & 0.412036 & 6740 ± 305 & 6845 & 1.44 ± 0.27 & 0.64 & 29.59 ± 3.80 & 32.68 & -- & $-$0.70 & 3.39 ± 0.15 \\
UV Vir & 377172421 & 0.56 ± 0.05 & 0.587065 & 0.586541 & 7550 ± 130 & 6714 & 1.75 ± 0.28 & 0.66 & 38.59 ± 6.88 & 49.96 & $-$1.10 ± 0.10 & $-$0.41 & 3.56 ± 0.11 \\
UZ CVn & 376689735 & 0.51 ± 0.03 & 0.697793 & 0.696838 & 6329 ± 58 & 6599 & -- & 0.66 & -- & 56.78 & $-$2.22 ± 0.09 & $-$0.85 & -- \\
SV Hya & 453469791 & 1.21 ± 0.05 & 0.478527 & 0.478892 & 6744 ± 324 & 6850 & 1.44 ± 0.28 & 0.71 & 31.79 ± 3.08 & 51.43 & -- & -- & 3.36 ± 0.15 \\
\bottomrule
\end{tabularx}
\end{table}

\begin{table}[H]\ContinuedFloat\small
\caption{\textit{Cont.}}
\label{tab:rrab_parameters}
\begin{tabularx}{\textwidth}{lLCCCCCCCcCcCC}
\toprule
\textbf{ID} & \textbf{TESS ID} & \boldmath\textbf{Plx$_{\rm lit}$} & \textbf{Period} & \textbf{Period LS} & \boldmath\textbf{T$_{\rm eff_{lit}}$} & \boldmath\textbf{T$_{\rm eff_{ANN}}$} & \boldmath\textbf{M$_{\rm lit}$} & \boldmath\textbf{M$_{\rm ANN}$} & \boldmath\textbf{Lum$_{\rm lit}$} & \boldmath\textbf{Lum$_{\rm ANN}$ }& \boldmath\textbf{[Fe/H]$_{\rm lit}$} & \boldmath\textbf{[Fe/H]$_{\rm ANN}$} & \boldmath\textbf{logg$_{\rm lit}$} \\
 &  & \textbf{('') }& \textbf{(d)} & \textbf{(d)} & \textbf{(K)} & \textbf{(K)} & \boldmath\textbf{(M$_{\odot}$)} & \boldmath\textbf{(M$_{\odot}$)} & \boldmath\textbf{(L$_{\odot}$)} & \boldmath\textbf{(L$_{\odot}$)} & \textbf{(dex)} & \textbf{(dex)} & \textbf{(dex)}\\
\midrule
UY Boo & 458457951 & 0.62 ± 0.05 & 0.650800 & 0.651139 & 6433 ± 292 & 6581 & -- & 0.77 & -- & 69.01 & -- & -0.73 & -- \\
RS Boo & 409373422 & 1.36 ± 0.04 & 0.377365 & 0.377471 & 6610 ± 132 & 6719 & 1.38 ± 0.22 & 0.61 & 24.05 ± 1.74 & 29.52 & $-$0.12 ± 0.10 & $-$0.21 & 3.43 ± 0.09 \\
V413 CrA & 253708643 & 1.13 ± 0.05 & 0.589343 & 0.589858 & 5945 ± 47 & 6521 & -- & 0.60 & -- & 50.39 & $-$0.94 ± 0.05 & $-$0.15 & -- \\
V4424 Sgr & 271404999 & 1.70 ± 0.04 & 0.424503 & 0.424840 & 6240 ± 181 & 6511 & 1.21 ± 0.18 & 0.55 & 25.46 ± 1.63 & 24.36 & -- & $-$0.00 & 3.25 ± 0.10 \\
TV Lib & 79403057 & 0.79 ± 0.04 & 0.269624 & 0.269903 & 6620 ± 132 & 6897 & 1.39 ± 0.23 & 0.64 & 17.68 ± 1.81 & 16.29 & $-$0.43 ± 0.10 & $-$0.28 & 3.57 ± 0.10 \\
BT Aqr & 248815852 & 0.56 ± 0.05 & 0.406357 & 0.406222 & 6700 ± 268 & 6681 & 1.42 ± 0.27 & 0.62 & 25.69 ± 4.50 & 37.35 & -- & $-$0.55 & 3.44 ± 0.15 \\
AA Aql & 248097218 & 0.68 ± 0.05 & 0.361768 & 0.361829 & 6550 ± 146 & 6915 & 1.35 ± 0.24 & 0.65 & 27.71 ± 4.25 & 30.46 & $-$0.32 ± 0.10 & $-$0.63 & 3.35 ± 0.12 \\
V456 Ser & 46253107 & 0.22 ± 0.04 & 0.517558 & 0.517328 & 6600 ± 40 & 6084 & -- & 0.76 & -- & 151.05 & $-$2.64 ± 0.17 & $-$0.66 & -- \\
VY Ser & 371355048 & 1.32 ± 0.07 & 0.714096 & 0.712959 & 6055 ± 50 & 6448 & -- & 0.70 & -- & 64.02 & $-$1.82 ± 0.05 & $-$0.36 & -- \\
V341 Aql & 375822497 & 0.84 ± 0.05 & 0.578042 & 0.579124 & 6504 ± 272 & 6666 & -- & 0.73 & -- & 63.17 & -- & $-$1.63 & -- \\
AT Ser & 264584937 & 0.58 ± 0.05 & 0.746609 & 0.745240 & 6591 ± 249 & 6436 & -- & 0.75 & -- & 73.06 & -- & $-$0.47 & -- \\
DX Del & 85211589 & 1.68 ± 0.03 & 0.472611 & 0.472674 & 6454 ± 118 & 6534 & 1.31 ± 0.21 & 0.60 & 32.36 ± 1.84 & 36.11 & $-$0.14 ± 0.06 & $-$0.16 & 3.24 ± 0.09 \\
VX Her & 356085581 & 0.98 ± 0.06 & 0.455356 & 0.456175 & 6369 ± 135 & 6786 & 1.27 ± 0.20 & 0.74 & 37.00 ± 4.74 & 46.23 & $-$1.45 ± 0.07 & -- & 3.14 ± 0.11 \\
BN Vul & 111443222 & 1.40 ± 0.03 & 0.594132 & 0.593941 & 6318 ± 222 & 6749 & -- & 0.85 & -- & 143.31 & -- & $-$0.86 & -- \\
TW Her & 10581605 & 0.86 ± 0.02 & 0.399596 & 0.400107 & 7465 ± 147 & 6836 & 1.71 ± 0.29 & 0.66 & 29.37 ± 1.94 & 36.89 & $-$0.35 ± 0.10 & $-$1.14 & 3.65 ± 0.09 \\
CN Lyr & 317170031 & 1.11 ± 0.03 & 0.411382 & 0.411798 & 6355 ± 106 & 6393 & 1.26 ± 0.19 & 0.54 & 22.33 ± 1.79 & 24.38 & $-$0.04 ± 0.10 & $-$0.01 & 3.36 ± 0.09 \\
VZ Her & 85516380 & 0.63 ± 0.02 & 0.440360 & 0.440466 & 5732 ± 153 & 6870 & -- & 0.71 & -- & 42.43 & $-$1.71 ± 0.01 & -- & -- \\
FN Lyr & 158553034 & 0.30 ± 0.03 & 0.527390 & 0.527304 & 6745 ± 410 & 6735 & -- & 0.78 & -- & 77.00 & -- & $-$1.09 & -- \\
SW For & 175492625 & 0.36 ± 0.02 & 0.803760 & 0.803760 & 6389 ± 176 & 6426 & -- & 0.67 & -- & 59.24 & -- & $-$0.26 & -- \\
U Pic & 259590223 & 0.73 ± 0.02 & 0.440373 & 0.440466 & 6642 ± 106 & 6718 & 1.40 ± 0.23 & 0.63 & 30.57 ± 2.50 & 35.96 & $-$0.69 ± 0.08 & $-$0.69 & 3.34 ± 0.09 \\
CD Vel & 34069197 & 0.57 ± 0.03 & 0.573516 & 0.572059 & 7099 ± 261 & 6558 & 1.58 ± 0.27 & 0.80 & 54.15 ± 6.25 & 75.49 & -- & $-$0.74 & 3.26 ± 0.12 \\
ST Pic & 150166721 & 2.00 ± 0.02 & 0.485749 & 0.486130 & 6549 ± 182 & 6422 & 1.35 ± 0.22 & 0.56 & 33.05 ± 1.70 & 28.96 & -- & $-$0.06 & 3.27 ± 0.10 \\
AE Tuc & 234507163 & 0.63 ± 0.02 & 0.414528 & 0.414589 & 6274 ± 116 & 6826 & 1.22 ± 0.17 & 0.64 & 20.32 ± 1.79 & 33.14 & $-$0.45 ± 0.07 & $-$0.78 & 3.36 ± 0.09 \\
W Tuc & 234518883 & 0.57 ± 0.03 & 0.642247 & 0.643122 & 6042 ± 106 & 6655 & -- & 0.72 & -- & 66.97 & $-$1.31 ± 0.08 & $-$1.18 & -- \\
BI Cen & 267930751 & 0.69 ± 0.03 & 0.453193 & 0.452948 & 7085 ± 341 & 6765 & 1.57 ± 0.28 & 0.86 & 47.25 ± 5.75 & 106.42 & -- & $-$0.82 & 3.32 ± 0.14 \\
YY Tuc & 220512467 & 0.43 ± 0.03 & 0.634857 & 0.634866 & 6350 ± 233 & 6595 & -- & 0.77 & -- & 69.81 & -- & $-$1.37 & -- \\
RV Phe & 425863844 & 0.52 ± 0.04 & 0.596419 & 0.596355 & 6471 ± 195 & 6504 & -- & 0.68 & -- & 56.92 & -- & $-$0.39 & -- \\
TY Aps & 258812822 & 0.62 ± 0.03 & 0.501720 & 0.501070 & 6588 ± 283 & 6749 & -- & 0.68 & -- & 42.88 & -- & -- & -- \\
EX Aps & 294832702 & 0.58 ± 0.03 & 0.471799 & 0.472235 & 6738 ± 297 & 6741 & 1.44 ± 0.27 & 0.64 & 29.83 ± 3.51 & 41.24 & -- & $-$0.87 & 3.39 ± 0.14 \\
V Ind & 126910093 & 1.50 ± 0.04 & 0.479594 & 0.479773 & 6551 ± 105 & 6742 & 1.36 ± 0.21 & 0.73 & 33.35 ± 2.39 & 44.11 & $-$1.32 ± 0.03 & -- & 3.27 ± 0.08 \\
MS Ara & 337440887 & 0.55 ± 0.04 & 0.524988 & 0.525991 & 6735 ± 254 & 6503 & 1.44 ± 0.25 & 0.81 & 45.39 ± 6.95 & 56.61 & -- & $-$0.69 & 3.21 ± 0.13 \\
CD-45 13628 & 129081118 & 1.26 ± 0.04 & 0.510247 & 0.510016 & 6133 ± 115 & 6305 & 1.16 ± 0.16 & 0.56 & 28.45 ± 2.40 & 30.13 & $-$0.35 ± 0.08 & $-$0.08 & 3.15 ± 0.09 \\
CD-42 14707 & 129112145 & 0.99 ± 0.27 & 0.551264 & 0.551323 & -- & 6649 & -- & 0.61 & -- & 44.16 & -- & $-$0.40 & -- \\
CD-48 13332 & 166463208 & 0.40 ± 0.04 & 0.579506 & 0.578464 & 6651 ± 349 & 6711 & -- & 0.67 & -- & 45.00 & -- & $-$1.79 & -- \\
GI Psc & 406413012 & 0.42 ± 0.05 & 0.734199 & 0.736294 & 6487 ± 262 & 6477 & -- & 0.71 & -- & 59.77 & -- & $-$0.34 & -- \\
\bottomrule
\end{tabularx}
   
\begin{adjustwidth}{+\extralength}{0cm}
 \footnotesize{\noindent{Notes:} 
 ID is the variable star identifier. TESS ID is the TESS Input Catalog identifier. Plx is the parallax in arcseconds. Period is the pulsation period in days (from the literature), and~Period LS is the period determined from the Lomb--Scargle periodogram applied to TESS light curves. Columns with subscript `lit' show values from the literature with their uncertainties when available. Columns with the subscript `ANN' show values derived from our ANN.}
\end{adjustwidth}
\end{table}
\finishlandscape
While the temperature and luminosity predictions show reasonable agreement with literature values, mass estimates from the ANN are systematically lower by approximately $0.74\, M_{\odot}$. From~stellar evolution, RR Lyrae stars are known to have masses in the range $0.5$--$0.8\, M_{\odot}$. In~contrast, many literature values (from the TIC catalogue) report anomalously high masses (1.1--1.7\,$M_{\odot}$), which is likely due to the use of isochrone or SED fitting methods that may not have accounted for the evolved, low-mass RR Lyrae~stars.

The relatively large spread in metallicity ($\sigma = 0.83$\, dex) and mean offset of $0.27$\, dex likely arises from degeneracies in light curve morphology, where variations in metallicity can mimic the effects of other parameters. Furthermore, literature metallicities are derived from diverse sources and methods, which introduces inconsistencies when used as a reference~benchmark.

Overall, the~ANN method demonstrates reasonable 
 predictive power for effective temperature and luminosity, while systematic shifts in mass and scatter in metallicity reflect a combination of astrophysical degeneracies and uncertainties in the literature values. This supports our approach of deriving fundamental stellar parameters directly from time-series photometry while also highlighting the need for caution when comparing to literature sources that may not be tailored for RR Lyrae-type~variables.

\subsection{Period--Luminosity--Metallicity (PLZ) Relation}

Using the stellar parameters inferred from the TESS RRab light curves via the trained ANN, we constructed a new empirical period--luminosity--metallicity (PLZ) relation. \textls[-25]{We fitted the logarithm of the luminosity as a linear function of the pulsation period and metallicity:}
\begin{equation}
    \log(L/L_\odot) = a \log(P/{\rm days}) + b\,[\mathrm{Fe/H}] + c,
\end{equation}
\textls[-25]{where $P$ is the pulsation period in days, and~$[\mathrm{Fe/H}]$ is the derived metallicity from the ANN-inferred $Z$ values. The~coefficients $a$, $b$, and~$c$ were determined using a least-squares fitting.}
\begin{equation}
    a = {1.458 \pm 0.028}, \qquad b = {-0.068 \pm 0.007}, \qquad c = {2.040 \pm 0.007}.
\end{equation}

\begin{figure}[H]
    \includegraphics[width=0.55\textwidth]{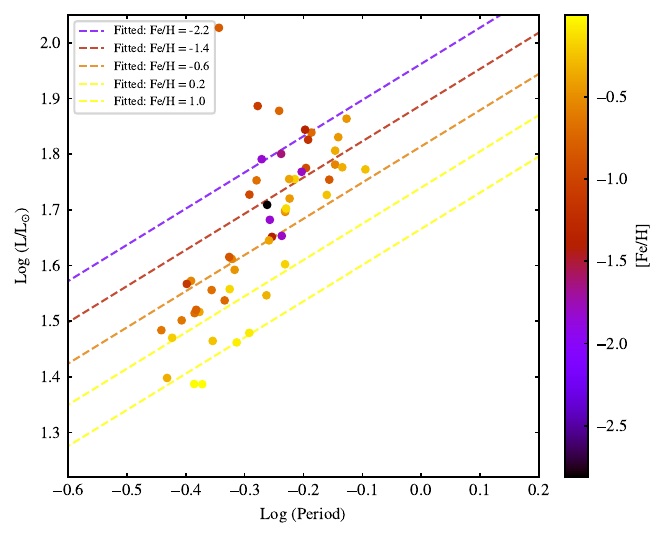}
    \caption{Period--luminosity--metallicity (PLZ) relation derived from ANN-inferred stellar parameters. The~background shows the fitted plane, while the scatter points represent individual stars color-coded by metallicity {[Fe/H].}} 
    \label{fig:PLZ_relation}
\end{figure}

Figure~\ref{fig:PLZ_relation} presents a plot of the ANN predicted luminosities as a function of period and metallicity, which was overlaid with the fitted PLZ plane. The~scatter of individual stars around the plane is shown as a color-coded scatter plot with~the color scale representing [Fe/H]. Stars with lower metallicity (bluer colors) lie systematically higher in luminosity at fixed period, which is consistent with theoretical~expectations.

\unskip

\section{Conclusions}

In this study, we developed and validated an artificial neural network (ANN)-based framework to infer the physical parameters of fundamental-mode RR Lyrae (RRab) stars directly from their light curves. Using a synthetic grid of pulsation models, we trained a feedforward ANN to learn the inverse mapping from $I$ band light-curve morphology to stellar parameters: mass ($M$), luminosity ($\log L$), effective temperature ($\log T_{\rm eff}$), and~metallicity ($Z$). The~network achieved high accuracy in recovering these parameters in synthetic self-inversion tests, demonstrating strong internal consistency and robustness across a wide parameter~space.

The self-inversion test, conducted on a hold-out test set of 1,715 models not seen during training, showed good agreement between predicted and true values, particularly for $\log T_{\rm eff}$ and $\log L$. However, the~performance was comparatively poorer for $M$ and $Z$, which was likely due to degeneracies in the light-curve morphology, where multiple combinations of parameters can lead to similar pulsation~features.

We then applied the trained parameter estimator model to observed RRab stars from the \textit{TESS} mission. After~preprocessing the light curves, including flux-to-magnitude conversion, phase folding, Fourier smoothing, extinction correction, and~distance modulus calibration, we extracted stellar parameters from the light curves of individual stars. The~values of $M$, $\log L$, $\log T_{\rm eff}$, and~$Z$, inferred using the model, were used to compute surface gravity ($\log g$) and iron abundance $[\mathrm{Fe/H}]$, providing a complete physical characterization based solely on photometric~data.

We compared the ANN-inferred values with those available in the literature (TIC v8.2 catalog). While the ANN-predicted values for $\log T_{\rm eff}$ and $\log L$ generally aligned well with the catalog values, the~agreement was notably worse for mass and metallicity. A~significant fraction of the TIC-inferred masses for RRab stars was found to lie in the range $1.1$–$1.7\, M_\odot$, which is inconsistent with established evolutionary models for RR Lyrae stars that predict a mass range of $0.5$–$0.8\, M_\odot$. These discrepancies suggest that catalog masses—often derived from isochrone or SED fitting—may not be reliable for evolved horizontal-branch stars like RR Lyraes. Consequently, the~differences are more likely due to limitations in the literature data rather than overfitting or failure of the ANN~model.


Furthermore, a~period–luminosity–metallicity (PLZ) relation was derived using the ANN-predicted parameters. While the PLZ relation is physically plausible, its interpretation is naturally constrained by the accuracy of the inferred stellar~properties.

This work demonstrates that an ANN-based inversion of RRab light curves offers an alternative to traditional model fitting and spectroscopic analysis. In the~future, the~method can be extended to include multi-band light curves, overtone RRc stars, and~different classes of pulsating variable stars like Cepheids and BL Herculis-type variables. This work is the first step and can provide a fast way to estimate global physical parameters when trained with an extended parameter space of RRab theoretical models.
Uncertainty-aware architectures and larger, better-constrained training sets will further improve the robustness and applicability of such data-driven frameworks in time-domain stellar~astrophysics.

\vspace{6pt} 






\authorcontributions{Conceptualization, H.P.S. and N.K.; 
methodology, N.K., H.P.S., O.M. and~S.J.; 
software, N.K.; 
validation, N.K.; 
formal analysis, N.K.; 
investigation, N.K.; 
resources, H.P.S. and S.J.; 
data curation, N.K.; 
writing---original draft preparation, N.K.; 
writing---review and editing, H.P.S., O.M., S.J., K.T., P.P. and~A.B.; 
visualization, N.K.; 
supervision, H.P.S.; 
project administration, H.P.S.; 
All authors have read and agreed to the published version of the~manuscript.
}

\funding{This research was funded by the BRICS STI Framework Programme grant
``Search and Follow-up Studies of Time-domain Astronomical Sources using Sky Surveys, BRICS Telescopes and Artificial Intelligence'' (SAPTARISI). H.P.S. and S.J. were supported by the Department of Science \& Technology (DST) through grant number DST/ICD/BRICS/Call-5/SAPTARISI/2023(G). O.M. was supported by the Ministry of Science and Higher Education of the Russian Federation, according to the research project 13.2251.21.0177 (075-15-2022-1228). K.T. was supported by the National Natural Science Foundation of China (NSFC) under grant No. 12261141689.}

\institutionalreview{Not~applicable.}

\dataavailability{The data will be made available on request.}

\acknowledgments{N.K. acknowledges the use of High Performance Cluster facility, Pegasus of IUCAA, Pune for providing the computational~resources.}

\conflictsofinterest{The authors declare no conflicts of~interest.} 



\abbreviations{Abbreviations}{
The following abbreviations are used in this manuscript:
\\

\noindent 
\begin{tabular}{@{}ll}
ANN & artificial neural network\\
PLZ & period--luminosity--metallicity (relation)\\
TESS & Transiting Exoplanet Survey Satellite
\end{tabular}
}

\printendnotes[custom]
\reftitle{References}

\PublishersNote{}


\end{document}